\documentclass[a4paper,10pt]{article}
\usepackage{amstext}
\usepackage{graphicx}

\title{Casimir-Polder forces from density matrix formalism}
\author{T.N.C. Mendes\footnote{tarciro@if.ufrj.br}$\;$ and C. Farina\footnote{farina@if.ufrj.br} \\
\textit{Instituto de F\'isica - UFRJ, Brazil}}
\date{December 21, 2005}

\begin{document}

\maketitle

\begin{abstract}
We use the density matrix formalism in order to calculate the energy level shifts, in second order on interaction, of an atom in the presence of a perfectly conducting wall in the dipole approximation. The thermal corrections are also examined when $\hbar \omega_0/k_B T = k_0 \lambda_T \gg 1$, where $\mbox{$\omega_0=k_0 c$}$ is the dominant transition frequency of the atom and $\lambda_T$ is the thermal length. When the distance $z$ between the atom and the wall is larger than $\lambda_T$ we find the well known result obtained from Lifshitz's formula, whose leading term is proportional to temperature and is independent of $c$, $\hbar$ and $k_0$. In the short distance limit, when $z\ll\lambda_T$, only very small corrections to the leading vacuum term occur. 
%
%We find a stable equilibrium point at $z\approx 0.5\lambda_T$ considering only the thermal %corrections (without vacuum term), which is a curious thing. 
%
We also show, for all distance regimes, that the main thermal corrections are independent of $k_0$ (dispersion is not important) and dependent of $c$, which means that there is not a non-retarded regime for the thermal contributions.
\end{abstract}
%
%%%%%%%%%%%%%%%%%%%%%%%%%%%%%%%%%%%%%%%%%%%%%%%%%%%%%%%%%%%%%%%%%%%%%%%%%%%%%%%%%%%%%%%%%%%%%%%%%%%%%%%%%
%%%%%%%%%%%%%%%%%%%%%%%%%%%%%%%%%%%%%%%%%%%%%%%%%%%%%%%%%%%%%%%%%%%%%%%%%%%%%%%%%%%%%%%%%%%%%%%%%%%%%%%%%
%
\section{Introduction}

In 1948, Casimir and Polder \cite{CasiPolder} considered for the first time the influence 
of retardation effects on the van der Waals forces between two atoms as well as on the 
force between an atom and a perfectly conducting wall. The non-retarded dispersive van der Walls force between two neutral, but polarizable atoms, had been explained previously by London in 1930 \cite{London}. Since Casimir and Polder's paper, forces between atoms or molecules and any kind of walls are usually called \textit{Casimir-Polder forces}. These authors were motivated by experiments performed by Verwey and Overbeek \cite{Overbeek} with colloidal suspensions which showed that London's interaction (that falls as $1/r^6$) was not correct for large distances, where the finite velocity of light should be taken into account. 
They showed that in the retarded limit the interaction falls as $1/r^7$ for two atoms (in contrast to London's result) and as $1/r^4$ for an atom and a perfectly conducting wall, in contrast with the short distance limit (that falls as $1/r^3$). 
%
%which can be easily calculated by the method of images \cite{CCT_QM}. 
%
Casimir and Polder obtained their results after lengthy calculations by using perturbative methods in QED \cite{CasiPolder}. However, following a Bohr's suggestion, Casimir rederived the results obtained  with Polder in a much simpler way, by computing the shift in the electromagnetic zero-point energy caused by the presence of the atoms and the walls \cite{Casimir1949}. A month later, Casimir presented his seminal paper on the attraction between two parallel conducting plates which gave rise to the famous effect that since then bears his name \cite{Casimir1948}.

In 1956, Lifshitz and collaborators developed a general theory of van der Waals forces \cite{Lifshitz}. They derived a powerful expression for the force at finite temperature between two semi-infinite dispersive media characterized by well defined dielectric constants and separated by a slab of any other dispersive medium. They were able to derive and predict several results, like the variation of the thickness of thin superfluid helium films in a remarkable agreement with the experiments \cite{SabAnd}. They also showed that the Casimir-Polder force is a limiting case when one of the media is sufficiently dilute such that the force between the slabs may be obtained by direct integration of a single atom-wall interaction \cite{DLP}. 

Since then, many works have been done about this subject and the theory of van der Waals and Casimir-Polder interactions can be considered almost complete by now. Good reviews have been written on dispersive van der Waals interaction \cite{Milonni,Langbein,Margenau} and many  elaborated papers concerning level-shifts near surfaces have appeared, as 
for example, \cite{Meschede,HindsSandoghdar,Jhe,NhaJhe}, to mention just a few. It is worth 
mentioning that Casimir-Polder forces have been observed experimentally \cite{HindsetAl93}.
Higher multipole corrections \cite{Salam}, roughness \cite{PAm} and corrugation of the surfaces \cite{Emig}, and the influence of the Casimir-Polder interaction on Bose-Einstein condensates \cite{Vuletic} are some of the many branches of great activity on this subject nowadays, and the limiting cases of the Lifshitz formula are the starting point of the theoretical treatment of the majority of these problems.

Here we propose an alternative way of computing the Casimir-Polder forces as well as 
 van der Waals forces between single atoms which may be useful in the understanding of such interactions. The procedure to be presented is based on the density matrix formalism, from which we derive the master equation \cite{DDC1984} for a small system interacting with a large one. It can be applied for any state of the field (not only vacuum or thermal states). Besides, it can take into account magnetic interactions (which are not considered in the Lifshitz formula) in a natural way.

Consider a system $\mathcal S + \mathcal R$, where $\mathcal S$ is a small system and $\mathcal R$ a reservoir. \linebreak Starting from an interacting Hamiltonian of the form $V = -\sum_j S_j R_j$, where $S_j$ and $R_j$ are $\mathcal S$ and $\mathcal R$ observables, the master equation comes from the perturbative expansion of the density matrix of the total system $\mathcal S + \mathcal R$. Going on up to second order, one can take the reduced trace over the Hilbert space of $\mathcal R$ and do the Markovian approximation. This approximation is due to the existence of two very different time scales, namely: the correlation time of the fluctuations of the $\mathcal R$ variables and the characteristic time evolution of $\mathcal S$, which is too large compared to the former. The final equation describes the evolution of the density matrix of $\mathcal S$, made up of a free term plus a linear term that accounts for the coupling with $\mathcal R$ \cite{AtPhInt}. One may then calculate the level shift of $\mathcal S$ and write it in terms of the linear susceptibility and symmetric correlation function of both small system and reservoir. This procedure permits us to distinguish  the roles played by $\mathcal S$ and $\mathcal R$ to the interaction \cite{DDC1984}.

Here, the small system will be an electrically polarizable atom and the radiation field will be considered as the reservoir. In the dipole approximation, the Hamiltonian has the bilinear form required by the master equation. We will then calculate the interaction in both vacuum and thermal states of the field and investigate all possible distance regimes, defined by the main transition frequency of the atom and the thermal length ($\lambda_T = \hbar c/k_B T$).
%
%%%%%%%%%%%%%%%%%%%%%%%%%%%%%%%%%%%%%%%%%%%%%%%%%%%%%%%%%%%%%%%%%%%%%%%%%%%%%%%%%%%%%%%%%%%%%%%%%%%%%%%%%
%%%%%%%%%%%%%%%%%%%%%%%%%%%%%%%%%%%%%%%%%%%%%%%%%%%%%%%%%%%%%%%%%%%%%%%%%%%%%%%%%%%%%%%%%%%%%%%%%%%%%%%%%
%
\section{The density matrix formalism}

In this section we will give a quick derivation of the master equation and obtain the general expressions for the level shifts. For details see references \cite{DDC1984,AtPhInt,AtPhQED}.

Consider the time dependent perturbation: 
\begin{equation}
\label{Vtdep}
V(t)=-\sum_j S_j(t)R_j(t)
\end{equation}

The equation for the evolution of the density matrix will be:
\begin{equation}\label{rholambint}
\frac{d}{dt}\rho\left( t\right) =\frac{i}{\hbar}\sum_j \left[ S_j(t)R_j(t),\rho\left( t\right)\right] 
\end{equation}
Admitting that the density matrix of $\mathcal S + \mathcal R$ may be approximated by $\rho=\rho_S\otimes\rho_R$, where $\rho_{S}$ is the reduced matrix of $\mathcal S$ defined by $\rho_{S}=\textrm{Tr}_{R}\left[ \rho \right]$ ($\rho_R$ is defined similarly), integrating (\ref{rholambint}) between $t_0$ to $t$ and making the adiabatic approximation \linebreak ($t_0 \rightarrow -\infty$), one can show in first order on the interaction that the expectation value of an observable $S_j$ at time $t$ is given by:
\begin{eqnarray}
\label{Smean}
\langle S_j\left( t\right) \rangle &=& \langle S_j\rangle +\sum_{k}\int_{-\infty}^{\infty}\!\!\!\!\! d\tau\; \chi_{jk}^S\left( \tau\right) \langle R_k\left( t-\tau\right) \rangle 
\\
%%%%%%%%%%%%%%%%%%%%%%%%%%%%%%%%%%%%%%%%%%%%%%%%%%%%%%%%%%%%%%%%%%%%%%%%%%%%%%%%%%%%%%%%%%%%%%%%%%%%%%%%%%%%%%%%
%
\label{qui11}
\chi_{jk}^{S}\left( \tau\right) &=& \frac{i}{\hbar}\Theta\left( \tau\right) \langle\left[ S_j\left( 0\right) ,S_k\left( -\tau\right) \right] \rangle 
\end{eqnarray}
where $\chi_{jk}^S\left( \tau\right)$ is the linear susceptibility of $\mathcal S$ respect to observables $S_j$ and $S_k$ \cite{Kubo1966} and $\Theta\left(\tau\right)$ is the 
Heaviside step function. The brackets mean that the ensemble average is over the non-perturbed states and $\langle S_j\rangle$ is the mean value of $S_j$ at $t=-\infty$. In the deduction of last equations it was not necessary to approximate $\mathcal R$ by a reservoir: the only assumption is that  the total density matrix is a simple direct product of the density matrices of the subsystems.

Considering now $\mathcal R$ as a reservoir, one can choose $\langle R_j\rangle=0$ for all relevant times without loss of generality, which causes the vanish of the first order term above. Another consequence for considering $\mathcal R$ as a reservoir is that the second order correlations of $\mathcal R$ observables are very sharp in time (of order of the inverse of a cutoff frequency) compared to correlations of $\mathcal S$ observables (Markovian approximation). Then, we can write for the density matrix of $\mathcal S$:
\begin{eqnarray}
\label{eqmestra}
\frac{d\rho_{ab}^S\left( t\right) }{dt} &=& -\; i\omega_{ab}\rho_{ab}^S\left( t\right)+\sum_{c,d}{\mathcal J}_{abcd}\rho_{cd}^S\left( t\right)
\\
%%%%%%%%%%%%%%%%%%%%%%%%%%%%%%%%%%%%%%%%%%%%%%%%%%%%%%%%%%%%%%%%%%%%%%%%%%%%%%%%%%%%%%%%%%%
%
\label{txmestra}
{\mathcal J}_{abcd}&=&-\;\frac{1}{\hbar^2}\sum_{j,k} \int_0^\infty d\tau
%\left\lbrace
\Biggl\{g_{jk}\left( \tau\right) \left[ \delta_{bd} \sum_n S_{an}^j S_{nc}^k e^{-i\omega_{nc}\tau} - S_{ac}^k S_{db}^j e^{-i\omega_{ac}\tau}\right]
%\right.
%
\nonumber \\ 
&+&\ g_{kj}\left( -\tau\right) \left[ \delta_{ac} \sum_n S_{dn}^k S_{nb}^j e^{-i\omega_{dn}\tau} - S_{ac}^j S_{db}^k e^{-i\omega_{db}\tau}\right]\Biggr\} 
%\right\rbrace
\end{eqnarray}
where $g_{jk}\left( \tau\right) = {\textrm Tr}_R\left[\rho_R R_j\left( \tau\right) R_k\left( 0\right)\right]$, $\rho_{ab}^S = \langle a \vert \rho_S \vert b \rangle$, $S_{ab}^j = \langle a \vert S_j \vert b \rangle$, \linebreak $\hbar\omega_{ab}=E_a-E_b$ and $\vert a\rangle$ represents an eigenstate of energy of $\mathcal S$ with eigenvalue $E_a$. Equation (\ref{eqmestra}) is the master equation for $\mathcal S$ and gives all information about the system: the first term gives the free evolution and the second one accounts for the coupling with the reservoir. In order to obtain the level shifts it is necessary to consider only the non-diagonal terms which are coupled with themselves or, in other words, the terms in which the indexes are such that $c = a$ and $d = b$. Then, from eq.(\ref{txmestra}), one may write:
\begin{eqnarray}
\label{levelshift}
{\mathcal J}_{abab}=-\Gamma_{ab} - i\Delta_{ab} \ ; \ \ 
\Delta_{a}=\frac{1}{\hbar}\sum_{\mu} p_{\mu} \sum_{\nu} \sum_n {\mathcal P}\frac{\vert\langle\nu,n\vert V\vert \mu, a\rangle\vert^2}{E_\mu+E_a-E_\nu-E_n}
\end{eqnarray}
where $\Delta_{ab} = \Delta_{a} - \Delta_{b}$, $\Gamma_{ab}$ is a damping, $\mathcal P$ is the Cauchy's principal value, $p_\mu$ is the statistical weight of the state $\vert \mu \rangle$ which is an eigenstate of $\mathcal R$ with energy eigenvalue $E_\mu$ and $V$ is defined by eq.(\ref{Vtdep})%
\footnote{Throughout this paper the greek indexes refer to reservoir energy eigenstates and the latin indexes refer to system energy eigenstates.}%
.

Coming back to $g_{jk}\left( \tau\right)$, one may split it into its real and imaginary parts, which are related to the symmetric correlation functions and the linear susceptibilities. Then, we write $\Delta_a = \Delta_a^{fr} +
\Delta_a^{rr}$ such that
%
%equation (\ref{levelshift}) can be written as:
%
\begin{eqnarray}
\label{fr}
\hbar \Delta_a^{fr}\!\!\!\!&=&\!\!\!\!-\frac{1}{2}\sum_{j,k}\int_{-\infty}^{\infty}\frac{d\omega}{2\pi}\chi_{jk}^{\prime S,a}\left( \omega\right) C_{jk}^{R}\left( \omega\right)
\\
%%%%%%%%%%%%%%%%%%%%%%%%%%%%%%%%%%%%%%%%%%%%%%%%%%%%%%%%%%%%%%%%%%%%%%%%%%%%%%%%%%%%%%%%%%
%
\label{rr}
\hbar \Delta_a^{rr}\!\!\!\!&=&\!\!\!\!-\frac{1}{2}\sum_{j,k}\int_{-\infty}^{\infty}\frac{d\omega}{2\pi}\chi_{jk}^{\prime R}\left( \omega\right) C_{jk}^{S,a}\left( \omega\right)
\\
%%%%%%%%%%%%%%%%%%%%%%%%%%%%%%%%%%%%%%%%%%%%%%%%%%%%%%%%%%%%%%%%%%%%%%%%%%%%%%%%%%%%%%%%%%%%
%
C_{jk}^{S,a}\left( \omega\right) \!\!\!\!&=&\!\!\!\! \pi \sum_{n} S_{a n}^{j} S_{n a}^{k}\left[ \delta\left( \omega+\omega_{an}\right) +\delta\left( \omega-\omega_{an}\right) \right]
\nonumber \\
%%%%%%%%%%%%%%%%%%%%%%%%%%%%%%%%%%%%%%%%%%%%%%%%%%%%%%%%%%%%%%%%%%%%%%%%%%%%%%%%%%%%%%%%%%%
%
\chi_{jk}^{\prime S,a}\left( \omega\right)\!\!\!\! &=&\!\!\!\! -\frac{1}{\hbar}\sum_{n} S_{a n}^{j} S_{n a}^{k}\left[ {\mathcal P}\frac{1}{\omega_{a n}+\omega}+{\mathcal P}\frac{1}{\omega_{a n}-\omega}\right] 
\\
%%%%%%%%%%%%%%%%%%%%%%%%%%%%%%%%%%%%%%%%%%%%%%%%%%%%%%%%%%%%%%%%%%%%%%%%%%%%%%%%%%%%%%%%%%%%
%
C_{jk}^{R}\left( \omega\right)\!\!\! &=&\!\!\! \pi \sum_{\mu}p_{\mu}\sum_{\nu}R_{\mu\nu}^{j} R_{\nu\mu}^{k}\left[ \delta\left( \omega+\omega_{\mu\nu}\right) +\delta\left( \omega-\omega_{\mu\nu}\right) \right]
\nonumber \\
\chi_{jk}^{\prime R}\left( \omega\right)\!\!\! &=&\!\!\! -\frac{1}{\hbar}\sum_{\mu}p_{\mu}\sum_{\nu}R_{\mu\nu}^{j} R_{\nu\mu}^{k}\left[ {\mathcal P}\frac{1}{\omega_{\mu\nu}+\omega}+{\mathcal P}\frac{1}{\omega_{\mu\nu}-\omega}\right] 
\end{eqnarray}
where $\delta E_a = \hbar\Delta_a=\hbar\Delta_a^{fr}+\hbar\Delta_a^{rr}$ is the energy level shift of $\mathcal S$ in the state $\vert a \rangle$, $\chi_{jk}^{\prime R}\left( \omega\right)$ is the real part of the susceptibility of $\mathcal R$ and $C_{jk}^{R}\left( \omega\right)$ is its symmetric correlation function, $C_{jk}^{S,a}\left( \omega\right)$ and $\chi_{jk}^{\prime S,a}\left( \omega\right)$ are the equivalents of $C_{jk}^{R}\left( \omega\right)$ and $\chi_{jk}^{\prime R}\left( \omega\right)$ for $\mathcal S$ in the state $\vert a \rangle$  and $R_{\mu\nu}^{j} = \langle \mu \vert R_j \vert \nu \rangle$. In the present formalism, equation (\ref{fr}) gives the contribution due to the polarization of the system under fluctuations of reservoir ($fr$) and  equation (\ref{rr}) gives the reaction on the reservoir due to the fluctuations of the system ($rr$). These equations will be the starting point for all calculations.
%
%%%%%%%%%%%%%%%%%%%%%%%%%%%%%%%%%%%%%%%%%%%%%%%%%%%%%%%%%%%%%%%%%%%%%%%%%%%%%%%%%%%%%%%%%%%%%%%%%%%%%%%%%%%%%%%
%%%%%%%%%%%%%%%%%%%%%%%%%%%%%%%%%%%%%%%%%%%%%%%%%%%%%%%%%%%%%%%%%%%%%%%%%%%%%%%%%%%%%%%%%%%%%%%%%%%%%%%%%%%%%%%
%
\section{Force between an atom and a wall}

Let us consider the standard Casimir setup: two perfectly conducting plates of side $L$ separated by a distance $\ell$, with $L\gg \ell$, and located at $z = 0$ and $z = \ell$. Let us consider also that there exists an atom at a position $z$, with $0 < z < \ell$. For a wave-vector ${\bf k}$, the vector potential at the atom position is given by \cite{Barton1987}:
\begin{eqnarray}
{\bf A}_{{\bf k}_{_{||}}n}\left( {\bf r}_{_{||}}, z, t \right)\!\!\!\!\! &=&\!\!\!\!\! \left( \frac{2\pi\hbar }{c k \ell L^2}\right)^{1/2}\!\! \Bigg\lbrace a^{(1)}_{{\bf k}_{_{||}}n}\left( \hat{k}_{_{||}}\times \hat{z}\right) \sin \left(\frac{n\pi }{\ell}z\right) + 
\\ 
\!\!\!\!\!\!\! &a&\!\!\!\!\!\!\!^{(2)}_{{\bf k}_{_{||}}n}\left[ i\frac{n\pi}{k \ell}\hat{k}_{_{||}} \sin \left(\frac{n\pi}{\ell}z\right) - \hat{z} \frac{k_{_{||}}}{k} \cos \left(\frac{n\pi}{\ell}z\right) \right] \!\!\Bigg\rbrace  e^{i\left( {\bf k}_{_{||}}\cdot {\bf r}_{_{||}} -\omega t\right)}\! + h.c. 
\nonumber
\end{eqnarray}
\begin{eqnarray}
{\bf E}_{{\bf k}_{_{||}}n} &=& ik{\bf A}_{{\bf k}_{_{||}}n}\;\; \textrm{and} \;\; {\bf B}_{{\bf k}_{_{||}}n} = i{\bf k}\times{\bf A}_{{\bf k}_{_{||}}n} 
\\
%%%%%%%%%%%%%%%%%%%%%%%%%%%%%%%%%%%%%%%%%%%%%%%%%%%%%%%%%%%%%%%%%%%%%%%%%%%%%%%%%%%%%%%%%%%%%%%
%
\left[a^{(i)}_{{\bf k}_{_{||}}n},a^{(j)}_{{\bf k}^{\prime}_{_{||}}n^{\prime}}\right] &=& \left[a^{\dag (i)}_{{\bf k}_{_{||}}n},a^{\dag (j)}_{{\bf k}^{\prime}_{_{||}}n^{\prime}}\right] = 0\; ; \; \left[a^{(i)}_{{\bf k}_{_{||}}n},a^{\dag (j)}_{{\bf k}^{\prime}_{_{||}}n^{\prime}}\right] =\delta_{ij}\delta_{n n^{\prime}}\delta_{{\bf k}_{_{||}}{\bf k}^{\prime}_{_{||}}}
\nonumber 
\end{eqnarray}
where $\omega^2/c^2 = k^2 = k_{_{||}}^2 + \left( n\pi/\ell\right)^2$, ${\bf k}_{_{||}} = k_x \hat x + k_y \hat y$, ${\bf r}_{_{||}} = x \hat x + y \hat y$ and $n$ is a positive integer number.

For an electrically polarizable atom the interaction Hamiltonian is:
\begin{equation}
V({\bf x},t) = -{\bf d}\left( t\right) \cdot {\bf E}\left( {\bf x}, t\right) 
\end{equation}
where ${\bf d}\left( t\right)$ is the dipole moment of the atom induced by the field. Since this Hamiltonian is bilinear in the atom and field operators, one may use expressions (\ref{fr}) and (\ref{rr}) to calculate the interaction. Considering the atom as a two level isotropic system with transition frequency $\omega_0 = k_0 c$, the correlations and susceptibilities of the atom and the field will be given by:
\begin{eqnarray}\label{CSfs}
\chi^{\prime a}\left( \omega\right)\!\!\!\! &=&\!\!\!\! \chi_{jj}^{\prime S,a}\left( \omega\right) = \frac{\alpha_0 k_0 c}{2}{\mathcal P} \left[\frac{1}{k_0 c+\omega} + \frac{1}{k_0 c-\omega}\right] 
\nonumber \\
C^a\left( \omega\right)\!\!\!\! &=&\!\!\!\! C_{jj}^{S,a}\left( \omega\right) = \frac{\pi}{2}\hbar  k_0 c \alpha_0 \Bigl[ \delta\left( k_0 c  + \omega\right) + \delta\left( k_0 c - \omega\right)\Bigr] 
\\
%%%%%%%%%%%%%%%%%%%%%%%%%%%%%%%%%%%%%%%%%%%%%%%%%%%%%%%%%%%%%%%%%%%%%%%%%%%%%%%%%%%%%%%%%%%%%%%%%%%%%
%
\chi^{\prime R}_{{\bf k}_{_{||}}n}\left( \omega\right)\!\!\!\! &=&\!\!\!\! \frac{2 \pi c k}{L^2\ell} \left( 1 - \frac{n^2 \pi^2}{k^2 \ell^2} \cos \frac{2 n \pi}{\ell}z \right) {\mathcal P} \left[\frac{1}{kc + \omega} + \frac{1}{kc - \omega}\right] 
\nonumber \\
C^{R}_{{\bf k}_{_{||}}n}\left( \omega\right)\!\!\!\! &=&\!\!\!\! \frac{2 \pi^2 \hbar c k}{L^2\ell} \Bigl( 2\langle n_k\rangle+1\Bigr) \left( 1 - \frac{n^2 \pi^2}{k^2 \ell^2} \cos \frac{2 n \pi}{\ell}z\right) \Bigl[ \delta\left( k c  + \omega\right) + \delta\left( k c - \omega\right)\Bigr] \nonumber
\end{eqnarray}
In the last equation, $\alpha_0=2\vert {\mathbf d} \vert^2/3\hbar\omega_0$ is the static polarizability of the atom and $\langle n_k\rangle$ is the average number of photons for a given frequency and carries the information about the state of the field. Note that the susceptibility of the field is independent of its state and of $\hbar$, which makes it a completely classical quantity (this happens under any boundary condition and in a linear medium). 

Combining (\ref{CSfs}) and (\ref{fr}), summing over all possible modes and taking $\ell\rightarrow \infty$, one may write for the ($rr$) and ($fr$) contributions:
\begin{eqnarray}\label{MW}
\delta E^{rr}\left( z\right)\!\!\!\! &=&\!\!\!\! \delta E_0^{rr}\left( z\right) = \frac{\hbar c}{\pi}\int_0^\infty k^3 \alpha_{-}\left( k\right) G\left( 2 k z\right) dk \; ;
\nonumber \\
\delta E^{fr}\left( z\right)\!\!\!\! &=& \!\!\!\! \delta E_0^{fr}\left( z\right) + \delta E_{sf}^{fr}\left( z\right)
\; ; \;\;
\delta E_0^{fr}\left( z\right)= \frac{\hbar c}{\pi}\int_0^\infty k^3 \alpha_{+}\left( k\right) G\left( 2 k z\right) dk 
\nonumber \\
\delta E_{sf}^{fr}\left( z\right)\!\!\!\! &=&\!\!\!\! \frac{2 \hbar c}{\pi}\int_0^\infty k^3 \alpha_{+}\left( k\right) \langle n_k\rangle G\left( 2 k z\right) dk 
\\
\alpha_{\mp}\left( k\right)\!\!\!\! &=&\!\!\!\! \frac{\alpha_0 k_0}{2} {\mathcal P}\left( \frac{1}{k + k_0} \pm \frac{1}{k - k_0}\right) 
\; ; \;\;
G\left( x\right)=\frac{\textrm{sin}\; x}{x} + 2\frac{\textrm{cos}\; x}{x^2} - 2\frac{\textrm{sin}\; x}{x^3} \nonumber .
\end{eqnarray}

A first consequence of these results is that whenever the field is in the vacuum state, the ($rr$) contribution dominates at short distances and the ($fr$) contribution dominates at large distances. For short distances ($k_0 z \ll 1$), large values of $k$ give the dominant contributions to the integrals: in this limit, $\alpha_{-}\sim 2/k$ and $\alpha_{+}\sim -2k_0/k^2$, which makes the ($rr$) term more important. At large distances ($k_0 z \gg 1$), only small values of $k$ are significant: $\alpha_{-}\sim -2k/k_0^2$ and $\alpha_{+}\sim 2/k_0$, and the roles of ($fr$) and ($rr$) contributions are interchanged. As another consequence, we should expect weak corrections to the vacuum term for short distances and strong changes in the behavior of the interaction at large distances if $\langle n_k\rangle\neq 0$, since the ($fr$) term is the only one that contributes.

Focusing on the vacuum contribution, we may write for any regime:
\begin{eqnarray}\label{VCC}
\delta E_0\left( z\right)\!\!\!\! &=&\!\!\!\! \delta E_0^{fr}\left( z\right) + \delta E^{rr}\left( z\right) = \frac{\hbar c}{8 \pi} \frac{k_0 \alpha_0}{z^3} {\mathcal H}_0(x_0) \ 
\\
{\mathcal H}_0(x)\!\!\!\! &=&\!\!\!\!  \left( x^2 -2\right) {\mathcal F} \left( x\right) + 2 x {\mathcal G}\left( x\right) - x  
\nonumber \\
{\mathcal F} \left( x\right)\!\!\!\! &=&\!\!\!\! \textrm{Ci}\left( x\right) \sin x - \textrm{si}\left( x\right)\cos x \ \ ; \ \ {\mathcal G} \left( x\right) = \frac{d}{dx}{\mathcal F} \left( x\right) \ 
\nonumber \\
\textrm{Ci}\left( x\right)\!\!\!\! &=&\!\!\!\! -\int_x^\infty dt\; \frac{\cos t}{t}\;\;  \textrm{and}\;\; \textrm{si}\left( x\right) = -\int_x^\infty dt\; \frac{\sin t}{t} \nonumber 
\end{eqnarray}
where $x_0 = 2 k_0 z$.

%%%%%%%%%%%%%%   FIGURE  1  %%%%%%%%%%%%%%%%%
\begin{figure}[!h]
\begin{center}
\vskip -1.1 cm
\includegraphics[width=4.0in]{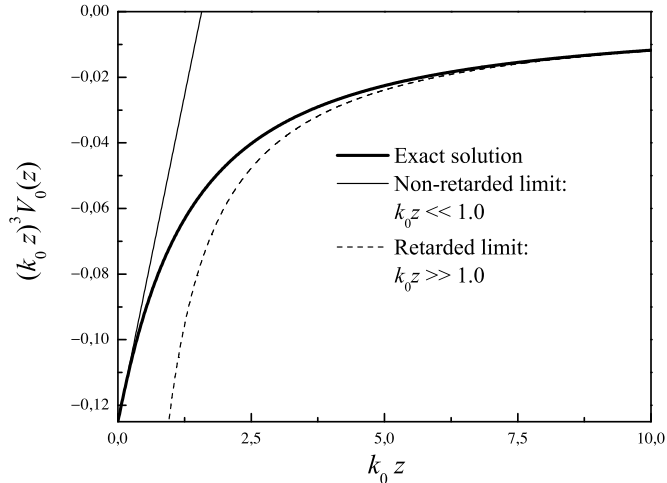}
\vskip -1.0 cm
%
%\footnotesize{
\caption{Exact solution (\ref{VCC}) and the asymptotic behaviors (\ref{NRL}) and  (\ref{CPR}) for $\left(k_0 z\right)^3 V_0(z)$, where $V_0(z)$ is the vacuum contribution, in units of $\hbar c \alpha_0 k_0^4$.}
%}
%
\label{V0}
\end{center}
\end{figure}
%
%%%%%%%%%%%%%%%%

In the non-retarded limit ($k_0 z \ll 1$), one may show that:
\begin{equation}\label{NRL}
V_0\left( z\right) = -\frac{\hbar \omega_0}{8} \frac{\alpha_0}{z^3} + {\mathcal O}\left( z^{-2}\right) 
\end{equation}
where $\omega_0 = k_0 c$. This result agrees with that obtained by the image method \cite{CCT_QM}.

In the retarded limit ($k_0 z \gg 1$), we have:
\begin{equation}
\label{CPR}
V_0\left( z\right) = -\frac{3}{8\pi}\hbar c \frac{\alpha_0}{z^4} + {\mathcal O}\left( z^{-6}\right)
\end{equation}
which is the well known Casimir-Polder result \cite{CasiPolder}. We show in Figure \ref{V0} the exact solution (\ref{VCC}) and the asymptotic behaviors (\ref{NRL}) and (\ref{CPR}).
%
%%%%%%%

%%%%%%%%%%%%%%%%%%%%%%%%%%%%%%%%%%%%%%%%%%%%%%%%%%%%%%%%%%%%%%%%%%%%%%%%%%%%%%%%%%%%%%%%%%%%%%%%%%%%%
%
\subsection{Thermal corrections}
Let us now consider the radiation field in a thermal state at a temperature $T$. 
Recently, thermal effects on intermolecular forces at non-zero temperature have 
been considered in the literature. For instance, limits of validity of the London-van der Waals potential between two atoms (assuming dilute media) have benn established \cite{Ninham}, a general analytic expression valid at any temperature $T$
have been derived for the interaction of two polarizable electric dipoles \cite{Goedecke}, or a Feinberg-Sucher theory \cite{FeinSuch} at finite temperature for two electrically-magnetically polarizable atoms in the non-dispersive regime
have been developed \cite{BartCPFS}. Here, we shall be concerned more specifically with the interaction between a polarizable atom and a perfectly conducting wall which is in thermal  equilibrium at temperature $T$. In this situation, the atom  interacts also with the corresponding thermal photons.

As one can see from equations (\ref{MW}), the contribution to the interaction is due only to the polarization of the atom by thermal fluctuations of the field:
\begin{equation}\label{MWT}
V_T(z) =\delta E_T^{fr}\left( z\right) = \frac{2\hbar c}{\pi}\int_0^\infty \frac{k^3 \alpha_{+}\left( k\right)}{ e^{\beta \hbar k c} - 1} G\left( 2 k z\right)dk
\end{equation}

\begin{figure}[htbp]
\begin{center}
\includegraphics[width=4.4in]{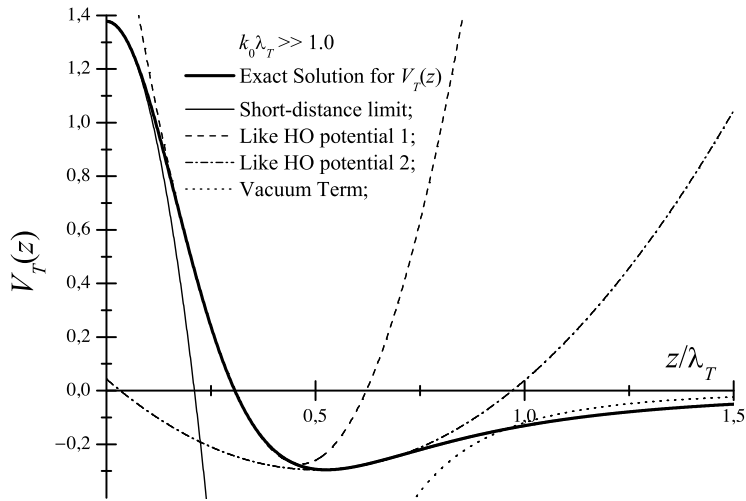}
\vskip -0.5 cm
\caption{Exact solution (\ref{VT}), the short distance limiting case (\ref{Vtnr}) for $0 < z/\lambda_T < 0.05$ and two like harmonic oscillator like potentials in the range $0.2 < z/\lambda_T < 0.45$ and $0.5 < z/\lambda_T < 0.75$ for $ V_T(z)$, which is the thermal correction to the interaction, in units of $\hbar c \alpha_0 /\lambda_T^4$. The vacuum term (\ref{CPR}) is also shown.}
\label{VTtot}
\end{center}
\end{figure}

In order to proceed it is convenient to introduce the quantity $\lambda_T = \beta \hbar c$. It defines the only relevant length scale for the thermal behavior of the interaction when $k_0 \lambda_T \gg 1$ (this is a necessary requirement to make the approximation of the atom by a two level system a good one) in the same way that $k_0$ defines the only length scale for the vacuum contribution. Beyond $\lambda_T$ thermal effects become more and more important. Then, equation (\ref{MWT}) may be written as:
\begin{eqnarray}
\label{VT}
V_T(z)\!\!\!\! &=&\!\!\!\! \frac{\hbar c}{8\pi}{\alpha_0 k_0 \over z^3}\Bigg[ \left(x_0^2 - 2\right) {\mathcal K}_0\left(\eta, x_0\right) +2 {\mathcal K}_1\left(\eta, x_0\right)-2 x_0 P\left(\eta\right)\Bigg]
\nonumber \\
%%%%%%%%%%%%%%%%%%%%%%%%%%%%%%%%%%%%%%%%%%%%%%%%%%%%%%%%%%%%%%%%%%%%%%%%%%%%
%
\label{K0exa}
{\mathcal K}_0\left(\eta, x_0\right)\!\!\!\! &=&\!\!\!\! {\textrm Im}\left[{\mathcal J}_0^{(+)}\left(\eta, x_0\right)-{\mathcal J}_0^{(-)}\left(\eta, x_0\right)\right] 
\nonumber \\
%%%%%%%%%%%%%%%%%%%%%%%%%%%%%%%%%%%%%%%%%%%%%%%%%%%%%%%%%%%%%%%%%%%%%%%%%%%
%
\label{K1exa}
{\mathcal K}_1\left(\eta, x_0\right)\!\!\!\! &=&\!\!\!\! -x_0{\textrm Re}\left[{\mathcal J}_0^{(+)}\left(\eta, x_0\right)+{\mathcal J}_0^{(-)}\left(\eta, x_0\right)\right] 
\\
%%%%%%%%%%%%%%%%%%%%%%%%%%%%%%%%%%%%%%%%%%%%%%%%%%%%%%%%%%%%%%%%%%%%%%%%%%%%%%
%
\label{J0-}
{\mathcal J}_0^{(-)}\left(\eta, x_0\right)\!\!\!\! &=&\!\!\!\!{i\pi e^{i x_0} \over e^{\eta x_0} -1}+\sum_{m = 1}^{\infty}e^{-\left(m\eta -i\right)x_0} E_1\left[-\left(m\eta -i\right)x_0\right]
\nonumber \\
%%%%%%%%%%%%%%%%%%%%%%%%%%%%%%%%%%%%%%%%%%%%%%%%%%%%%%%%%%%%%%%%%%%%%%%%
%
\label{J0+}
{\mathcal J}_0^{(+)}\left(\eta, x_0\right)\!\!\!\! &=&\!\!\!\!\sum_{m = 1}^{\infty}e^{\left(m\eta -i\right)x_0} E_1\left[\left(m\eta -i\right)x_0\right];\;\; P\left(\eta\right)=\sum_{m = 1}^{\infty}{1 \over 1+m^2\eta^2}
\nonumber
\end{eqnarray}
where $\eta x_0 = k_0 \lambda_T$, $x_0 = 2 k_0 z$ and $E_1\left(x\right)=\int_x^{\infty}dt\; e^{-t}/t$. For small values of $z/\lambda_T$, one can approximate the potential (\ref{VT}) by:
\begin{eqnarray}
\label{V_T(z)}
V_T(z)\!\!\!\! &=&\!\!\!\! \frac{\hbar c}{2 \pi}\frac{\alpha_0 k_0^2}{z^2}{\mathcal H}_T(x_0,\eta); \;\; Q\left( x\right)=\sum_{m = 2}^\infty \left(-1\right)^m\left(1 - {1 \over m}\right)\frac{\zeta\left(2 m\right)}{x^{2 m}}
\\
%%%%%%%%%%%%%%%%%%%%%%%%%%%%%%%%%%%%%%%%%%%%%%%%%%%%%%%%%%%%%%%%%%%%%%%%%%%%
%
\label{HT}
{\mathcal H}_T(x_0,\eta)\!\!\!\! &\simeq&\!\!\!\! Q\left( \eta\right) +%
\sum_{m = 2}^\infty \frac{(-1)^m x_0^{2 m -1}}{\left(2 m-1\right)!}\left[ {2\over x_0} - x_0 \left( 1 - {1 \over m}\right) \right] \sum_{j = m}^{N} \frac{\left( 2 j -1\right)!\zeta\left( 2 j\right) }{\left( \eta x_0\right)^{2 j} }
\nonumber
%%%%%%%%%%%%%%%%%%%%%%%%%%%%%%%%%%%%%%%%%%%%%%%%%%%%%%%%%%%%%%%%%%%%%%%
%
\end{eqnarray}
where $N \sim \eta x_0$ is an integer and $\zeta$ is the usual Riemann zeta function. The leading term of (\ref{V_T(z)}) is given by:
\begin{equation}\label{Vtnr}
V_{T}\left( z\right) \simeq C\left(T\right)-{\left(2 \pi\right)^5 \over 315}{\hbar c \alpha_0 \over \lambda_T^6} z^2 + {\mathcal O}\left(z^4\right)
\end{equation}
where $C\left(T\right)\simeq 1.38 \hbar c \alpha_0/\lambda_T^4$ 
 is independent of $z$ and, therefore, does not contribute to the interaction. This is a small repulsive
\footnote{This result does not contradict the one found in reference \cite{BartCPFS} for the thermal correction to the interaction between two atoms, $V_{T}^{AB}(r) = -C_{AB} T^6/r+\mathcal O\left(r\right)$ , since this last equation can not be used to derive (\ref{Vtnr}) by integration, because its validity is restricted to $1/k_0 \ll r \ll \lambda_T$, while (\ref{Vtnr}) takes into account contributions of all possible distances.}
correction to the vacuum term ($\vert V_T\vert \sim (z/ \lambda_T)^6 \vert V_0\vert$) and gives a very good approximation for the true $V_T(z)$ when $z \leq 0.05\lambda_T$, as one can see in Figure \ref{VTtot}: for $k_0\lambda_T \simeq 100$, which is the usual value for optical transition frequencies at room temperature, this means that $k_0 z \leq 5$; beyond this value, coincidently, the retarded effects become more relevant in the vacuum contribution (see Figure \ref{V0}). One may note also that (\ref{Vtnr}) is independent of $k_0$ and is $c$-dependent, which contrasts to (\ref{NRL}). This is an interesting result: at very short distances one could
naively expect that the effects of the finiteness of the velocity of light might be completely ignored, which would imply an interaction independent of $c$. Though it is the case for the  vacuum contribution, this does not occur for the thermal contribution: it is not possible to have an instantaneous field due to thermal fluctuations analogous to the Coulomb field at short distances.

%%%%%%%%%%%%%%%%%%%%%%%%%
%%%%%%%%%%%%%%%%%%%   FIGURE  2 CORRETA   %%%%%%%%%%%%%%%%%%%%%%%%%%%%%%%%%%%%%%%%%%%%%%%%%%%%%%%%%%%%%%%
\vskip -0.5 cm
\begin{figure}[htbp]
\begin{center}
\includegraphics[width=4.3in]{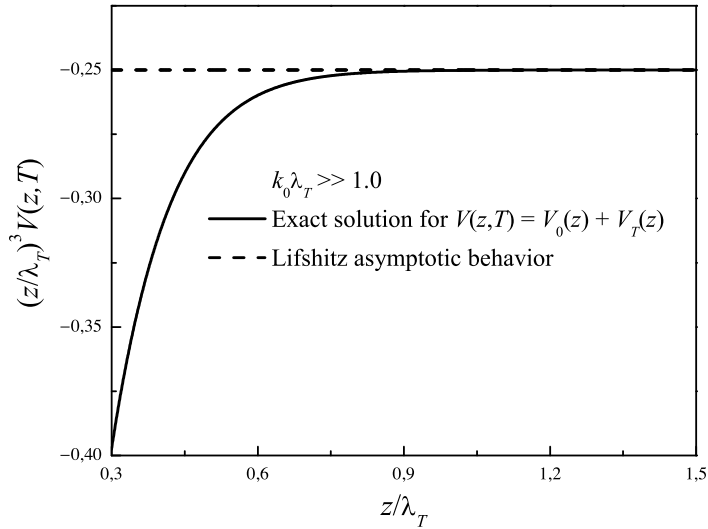}
\vskip -0.5 cm
\caption{Exact solution (\ref{VT}) plus (\ref{VCC}) and the asymptotic Lifshitz behavior (\ref{Lifshitz}) for $\left(z/\lambda_T\right)^3 V(z,T)$, where $V(z,T)$ is the total interaction, in units of $\hbar c \alpha_0 / \lambda_T^4$.}
\label{Vlarge}
\end{center}
\end{figure}
%%%%%%%%%%%%%%%%%%%%%%%%%%%%%%%%%%%%%%%%%%%%%%%%%%%%%

%%%%%%%%%%%%%%%%%%%%%%%%%%%%%%%%%%%%%%%%%%%%%%%%%%%%%%%%%%%%%%%%%%%

For distances $z \sim \lambda_T$ or larger, expression (\ref{V_T(z)}) is not convenient. In this case, one can approximate the potential written in (\ref{VT}) by:
\begin{eqnarray}\label{WT2}
V_T(z)\!\!\!\! &=&\!\!\!\! \frac{\hbar c}{2 \pi}\frac{\alpha_0 k_0}{z^3}{\mathcal L}_T(x_0,\eta) \ \ ; \ \ {\mathcal L}_T(x_0,\eta)\simeq -{1 \over x_0}P\left(\eta\right) + \left( 1 - {x_0^2 \over 2} \right)\times
\\
\!\!\!\!& \times &\!\!\!\!\sum_{j=1}^N\frac{1}{(\eta x_0)^{2 j + 1}}{\textrm I}m\left[ {\mathsf F}^{(2j)}\left(-\frac{i}{\eta} \right) \right] + x_0\sum_{j=1}^N\frac{1}{(\eta x_0)^{2 j}}{\textrm R}e\left[ {\mathsf F}^{(2j-1)}\left(-\frac{i}{\eta} \right) \right]
\nonumber
%%%%%%%%%%%%%%%%%%%%%%%%%%%%%%%%%%%%%%%%%%%%%%%%%%%%%%%%
%
\end{eqnarray}
where ${\mathsf F}^{(m)}\left(x\right)$ is the polygamma function of order $m$. Keeping the 
dominant terms of (\ref{WT2}) and summing them with the vacuum contribution, one may write for the asymptotic total interaction:
\begin{equation}\label{Lifshitz}
V(z,T)=V_0(z)+V_T(z)\simeq -\frac{k_B T}{4} \frac{\alpha_0}{z^3}
\end{equation}
which is the well known result obtained from Lifshitz formula \cite{Lifshitz,DLP}. Looking at Figure \ref{Vlarge} we see that the potential (\ref{Lifshitz}) is in excellent agreement with the exact solution for $z \sim \lambda_T$. For $z > \lambda_T$ the agreement is perfect (at room temperature this already occurs for $z > 7.6 \mathsf{\mu m}$). For comparison, one may note that the asymptotic behavior of the Casimir-Polder force given by (\ref{CPR}) fits the exact potential in a perfect way only for $z > 1.3\lambda_0$, where $\lambda_0 = 2\pi/k_0$.
%
%A curious thing about (\ref{Lifshitz}) is that it is a classical result, since it is not dependent on $\hbar$ or $c$: the thermal fluctuations of the field behave classically in this limit and the atom sees a classical field (however, the quantum nature still survives in $\alpha_0$).

%%%%%%%%%%%   FIGURE  3 CORRETA   %%%%%%%%%%%%
%
%%%%%%%%%%%%%%%%%%%%%
%%%%%%%%%%%%%%%%%%%%%%%%%%%%%%%%%%%%%%%%%%%%%%%%

In the intermediate distance regime  the thermal contribution to the interaction has a
 rich behavior. In Figure \ref{VTtot} we plot this contribution in the interval $0 < z/\lambda_T < 1.5$. As one can see, there would be a stable equilibrium point (along the ${\cal OZ}$ direction) at $z\simeq 0.52\lambda_T$ if this term were the only one contributing to the total interaction. For $z < 0.52\lambda_T$, the thermal contribution to the interaction between the atom and the wall is repulsive, and in the other interval $0.52 < z/\lambda_T < \infty$, the thermal contribution has an attractive character. In the range $0.2 < z/\lambda_T < 0.75$ one may fit at least two like harmonic oscilator potentials: in $0.2 < z/\lambda_T < 0.45$ and $0.5 < z/\lambda_T < 0.75$.

However, for the range $0.0 < z/\lambda_T < 0.75$, the vacuum contribution  dominates completely over the thermal contribution: despite the complex thermal contribution behavior,  this means that the total interaction is always attractive in this interval. For the other interval, $0.75 < z/\lambda_T < \infty$, the thermal term starts to compete with the vacuum term and becomes dominant from $z\simeq \lambda_T$ to $z = \infty$: but here, the thermal contribution is also attractive, so that the total potential $V(z,T) = V_0(z)+V_T(z)$ is again attractive.
In conclusion, we can state that, for all distance regimes, the force between a polarizable atom and a perfectly conducting wall is always attractive, which is a well established result.
%
%%%%%%%%%%%%%%%%%%%%%%%%%%%%%%%%%%%%%%%%%%%%%%%%%%%%%%%%%%%%%%%%%%%%%%%%%%%%%%%%%%%%%%%
%%%%%%%%%%%%%%%%%%%%%%%%%%%%%%%%%%%%%%%%%%%%%%%%%%%%%%%%%%%%%%%%%%%%%%%%%%%%%%%%%%%%%%%
%
\section{Conclusions}

In this paper we discussed the standard example of Casimir-Polder force: an electrically polarizable atom interacting with a perfectly conducting wall. We used the expressions for level shift of the atom (considered as a two level system) due to the field (considered as a reservoir) calculated from the master equation. This approach allowed us to split the level shift into two terms: the reservoir reaction ($rr$) term which dominates at short distances and is independent of the state of the field; the fluctuation of the reservoir ($fr$) term which is more important at large distances and carries the information about the field state. We found the exact behavior of the interaction in the vacuum state in the dipole approximation and the correct limits at short and large distances (non-retarded and retarded limits respectively).

We also considered the thermal state of the field. In this case we obtained the asymptotic behavior for distances $z >\lambda_T$ which gives a result in perfect agreement with that encountered in the literature \cite{Lifshitz,DLP}. We also found some results in the short distance limit ($z < \lambda_T$) which are small corrections to the vacuum terms: $\vert V_T\vert \sim (z/ \lambda_T)^6 \vert V_0\vert$. Considering only thermal corrections, we found a rich behavior for the interaction for $z < \lambda_T$: repulsive and attractive zones with one stable equilibrium point at $z\simeq 0.52\lambda_T$. In the short limit distance we found a thermal term depending on $c$, which is expected for $k_0 z \gg 1$ but not for $k_0 z \ll 1$, where the velocity of light should be (naively) ignored: we showed that it is not possible to talk about an instantaneous interaction like Coulomb field for short distances when thermal corrections are taken into account, though this surprisingly occurs at sufficiently large distances ($\lambda_T \ll z$). We think the formalism just presented may be useful for 
the computation of interatomic (intermolecular) interactions in many other situations, including magnetically polarizable atoms as well as other states of the radiation field. We hope to 
consider these problems in future works.
%
%%%%%%%%%%%%%%%%%%%%%%%%%%%%%%%%%%%%%%%%%%%%%%%%%%%%%%%%%%%%%%%%%%%%%%%%%%%%%%%%%%%%%%%%%%%%%%%%%%%%%%%%%%%%%%%%%
%%%%%%%%%%%%%%%%%%%%%%%%%%%%%%%%%%%%%%%%%%%%%%%%%%%%%%%%%%%%%%%%%%%%%%%%%%%%%%%%%%%%%%%%%%%%%%%%%%%%%%%%%%%%%%%%%
%
\vskip .5 cm
{\noindent\large\textbf{Acknowledgement}}
\vskip .3 cm
{\noindent The authors are grateful to Capes and CNPq for financial support. We are also indebted with the Second Referee for valuables suggestions}
%
%%%%%%%%%%%%%%%%%%%%%%%%%%%%%%%%%%%%%%%%%%%%%%%%%%%%%%%%%%%%%%%%%%%%%%%%%%%%%%%%%%%%%%%%%%%%%%%%%%%%%%%%%%%%
%%%%%%%%%%%%%%%%%%%%%%%%%%%%%%%%%%%%%%%%%%%%%%%%%%%%%%%%%%%%%%%%%%%%%%%%%%%%%%%%%%%%%%%%%%%%%%%%%%%%%%%%%%%%
%
\footnotesize

{

\end{document}
\begin{thebibliography}{99}

\bibitem {CasiPolder} Casimir H B G  and Polder D 1948 \textit{Phys. Rev.} {\bf 73} 360
%
\bibitem {London} London F 1930 \textit{Z. Physik} {\bf 63} 245
%
\bibitem{Overbeek}  Vervey E J W Overbeek J T G and van Nes K 1947 \textit{J. Phys. and Colloid Chem.} {\bf 51} 631
%
\bibitem{CCT_QM} Cohen-Tannoudji C Diu B and Lalo\"e F 1977 Quantum Mechanics Vol. 2 \textit{John Wiley and Sons Inc. New York} pp 1139-40
%
\bibitem{Casimir1949} Casimir H B G 1949 \textit{J. Chim. Phys.} {\bf 46} 407
%
\bibitem{Casimir1948} Casimir H B G 1948 \textit{Proc. K. Ned. Akad. Wet.} {\bf 51} 793
%
\bibitem{Lifshitz} Lifshitz E M 1956 \textit{Sov. Phys.} \textbf{JETP 2} 73
%
\bibitem{SabAnd} Sabisky E S and Anderson C H 1973 \textit{Phys. Rev. A} {\bf 7}(2) 790
%
\bibitem{DLP} Dzyaloshinskii I E Lifshitz E M and Pitaevskii L P 1961 \textit{Advan. Phys.} {\bf 10}(38) 165
%
%
\bibitem{Milonni} P.W. Milonni, {\it The Quantum Vacuum: An Introduction to Quantum Electrodynamics} (Academic, New York, 1994).
%
\bibitem{Langbein} Dieter Langbein, {\it theory of Van der Waals Attraction}, Springer Tracts in Modern Physics, Vol. {\bf 72} (Springer-Verlag, Berlin, 1974).
%
\bibitem{Margenau} H. Margenau and N.R. Kestner, {\it Theory of Intermolecular Forces} (Pergamon, New York, 1969).
%
\bibitem{Meschede} Meschede D Jhe W and Hinds E A 1990 {\it Phys. Rev. A} {\bf 41} 1587
%
\bibitem{HindsSandoghdar} Hinds E A and Sandoghdar V 1991 {\it Phys. Rev. A} {\bf 43} 398
%
\bibitem{Jhe} Jhe W 1991 {\it Phys. Rev. A} {\bf 43} 5795
%
\bibitem{NhaJhe} Nha H and Jhe W 1996 {\it Phys. Rev. A} {\bf 54} 3505
%
\bibitem{HindsetAl93} Sukenik C I Boshier M G Cho D  Sandoghdar V and Hinds E A 1993 
{\it Phys. Rev. Lett.} {\bf 70} 560
%
\bibitem{Salam} Salam A and Thirunamachandran 1996 \textit{J.Chem.Phys.} \textbf{104} 5094
%
\bibitem{PAm} Maia Neto P A Lambrecht A and Reynaud S 2005 \textit{Phys. Rev. A} \textbf{72} 012115
%
\bibitem{Emig} Emig T Hanke A Golestanian R and Kardar M 2003 \textit{Phys. Rev. A} \textbf{67} 022114
%
\bibitem{Vuletic} Lin Y J Teper I Ching C Vuletic V 2004 \textit{Phys. Rev. Lett} {\bf 92}(5) 050404
%
\bibitem{DDC1984} Dalibard J Dupon-Roc J and Cohen-Tannoudji C 1984 \textit{J. Physique} {\bf 45} 637
%
\bibitem{AtPhInt} Cohen-Tannoudji C Dupon-Roc J and Grynberg G 1992 Atom-Photon Interactions: Basic Processes and Applications \textit{John Wiley and Sons, Inc., New York} pp 262-321 
%
\bibitem{AtPhQED} Cohen-Tannoudji C Dupon-Roc J and Grynberg G 1989  \textit{John Wiley and Sons, Inc., New York} pp 352-56
%
\bibitem{Kubo1966} Kubo R 1966 \textit{Rep. Prog. Phys.} {\bf 29} 255
%
\bibitem{Barton1987} Barton G 1987 \textit{Proc. R. Soc. Lond. A} {\bf 410} 141
%
\bibitem{Ninham} Ninham B W and Daicic J 1998 {\it Phys. Rev. A} {\bf 57} 1870-80
%
\bibitem{Goedecke} Goedecke G H and Wood R C 1999 {\it Phys. Rev. A} {\bf 60} 2577-80
%
\bibitem{FeinSuch} Feinberg G and Sucher J 1970 {\it Phys. Rev. A} {\bf 2} (6) 2395
%
\bibitem{BartCPFS} Barton G 2001 \textit{Phys. Rev. A} {\bf 64} 032102
%
\end{thebibliography}
